\documentclass[superscriptaddress,twocolumn,prb,aps,10pt,longbibliography]{revtex4-1}

\usepackage{amsfonts,amssymb}
\usepackage[sumlimits,intlimits]{amsmath}
\usepackage{graphics}
\usepackage{graphicx}
\usepackage{mathrsfs}
\usepackage{textcomp}
\usepackage{verbatim}
\usepackage{bm}          % bold math letters

\newcommand{\be}{\begin{equation}}
\newcommand{\ee}{\end{equation}}
\newcommand{\lb}{\ell_B}
\newcommand{\Vt}{\widetilde{V}}
\newcommand{\e}{\varepsilon}
\newcommand{\rr}{\mathbf{r}}

\begin{document}

\title{Chemical potential and compressibility of quantum Hall bilayer excitons}

\author{Brian Skinner}
\affiliation{Materials Science Division, Argonne National Laboratory, Argonne IL, 60439, USA}
\affiliation{Massachusetts Institute of Technology, Cambridge, MA  02139 USA}\thanks{present address}

\date{\today}

\begin{abstract}

This paper considers a system of two parallel quantum Hall layers with total filling factor $0$ or $1$.  When the distance between the layers is small enough, electrons and holes in opposite layers form inter-layer excitons, which have a finite effective mass and interact via a dipole-dipole potential.  I present results for the chemical potential $\mu$ of the resulting bosonic system as a function of the exciton concentration $n$ and the interlayer separation $d$.  Both $\mu$ and the interlayer capacitance have an unusual nonmonotonic dependence on $d$, owing to the interplay between an increasing dipole moment and an increasing effective mass with increasing $d$.  A phase transition between  superfluid and Wigner crystal phases is shown to occur at  $d \propto n^{-1/10}$.  Results are derived first via simple intuitive arguments, and then verified with more careful analytic derivations and numeric calculations.

\end{abstract}

\maketitle

\section{Introduction} 

The idea of realizing Bose condensation or superfluidity in a solid state system goes back more than half a century.\citep{blatt_bose-einstein_1962, casella_possibility_1963, gergel_superfluidity_1968, gunton_condensation_1968, handel_van_1971,  keldysh_collective_1968}  The basic electronic excitations in solid state systems, electrons and holes, are fermionic, but these can be combined to form an electron-hole bound state, an exciton, that has bosonic statistics.  A collection of excitons in a semiconductor is therefore capable of assuming a Bose-Einstein condensate (BEC) or superfluid phase.  Such phases are difficult to realize in three-dimensional semiconductors, since the electrons and holes can recombine and thereby eliminate the exciton population.  But in two-dimensional (2D) systems the exciton population can be maintained by electrostatic gating or optical pumping, and indeed the first measurements of exciton condensation have been seen during the last fifteen years in two-dimensional semiconductors.\cite{butov_anomalous_1998, eisenstein_boseeinstein_2004, spielman_resonantly_2000,butov_macroscopically_2002}

\begin{figure}[t]
\centering
\includegraphics[width=0.45 \textwidth]{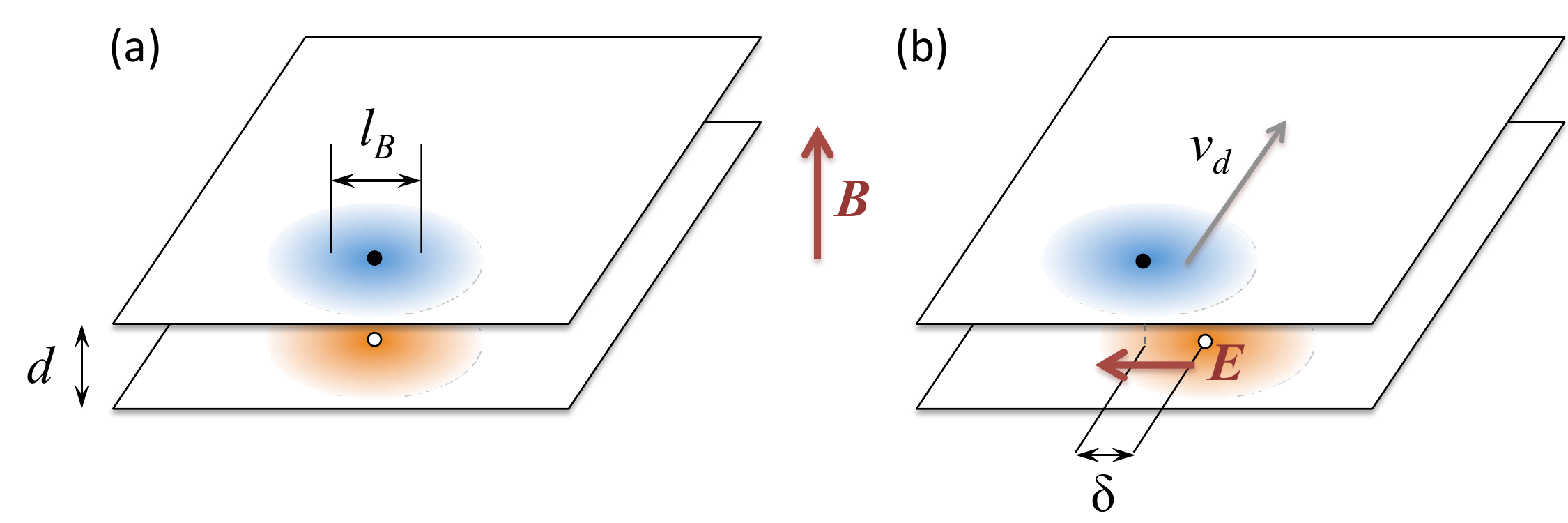}
\caption{(Color online) Schematic depiction of a quantum Hall bilayer exciton.  (a) The exciton comprises electron (blue area) and hole (orange area) wave packets, each with size $\sim \lb$, residing in opposite layers and bound by their mutual Coulomb attraction.  (b) The finite mass of the exciton arises from the relationship between the internal Coulomb energy of the exciton and its drift velocity.  When the electrons and hole are displaced laterally relative to each other by an amount $\delta$, the resulting crossed electric and magnetic fields ($\mathbf{E}$ and $\mathbf{B}$, respectively) produce a drift with velocity $v_d \propto \delta$. }
\label{fig:schematic}
\end{figure}

One particularly fruitful method for producing a stable excitonic condensate, as first suggested by Lozovik and Yudson,\cite{lozovik_feasibility_1975} is to spatially separate the electron and hole into parallel layers [as illustrated in Fig.\ \ref{fig:schematic}(a)].  In this spatially separated, bilayer configuration the electron and hole are prevented from recombining by the suppression of inter-layer tunneling.  However, if the interlayer spacing $d$ remains small enough, then the electron and hole can  still be bound together by their mutual Coulomb attraction to form a bosonic exciton.  Realization of exciton condensation was therefore largely enabled only by the development of sufficiently clean and sufficiently thin bilayer devices.  The ongoing, rapid development of ultra-clean, nanoscale 2D materials continues to provide new platforms and contexts for realizing bilayer exciton physics.\cite{zhang_excitonic_2008, seradjeh_exciton_2009,  rivera_observation_2015}

So far, most experimental studies of bilayer exciton condensates have sought to exploit the quantization of the electron and hole energies that arises in the presence of a large transverse magnetic field.\cite{lerner_two-dimensional_1981, butov_anomalous_1998, eisenstein_boseeinstein_2004, tiemann_exciton_2008, nandi_exciton_2012}  Such a magnetic field effectively quenches the electron kinetic energy by quantizing the electron and hole motion into Landau levels.  By thusly eliminating any competing Fermi energy, the electrons and holes can more readily bind together to form excitons.  
%The first evidence for bilayer exciton condensation has recently been seen in bilayers of GaAs, in the situation where each layer has filling factor $\nu$ close to $1/2$.

In this paper I consider the properties of a system of such quantum Hall bilayer excitons as a function of the exciton concentration $n$ and of the separation $d$ between layers.  I focus on the case where there is an equal concentration $n$ of electrons in the (say) top layer and holes in the bottom layer, and where the density is small enough that the corresponding filling factor $\nu = 2 \pi n \lb^2 < 1$.  Here, $\lb = \sqrt{\hbar c/eB}$ is the magnetic length (in CGS units, which are used throughout this paper); $\hbar$ is the reduced Planck constant, $c$ is the speed of light, $e$ is the electron charge, and $B$ is the magnetic field.  This description applies to either a pair of conventional semiconductor quantum wells with filling factors $\nu < 1$ and $1 - \nu$, respectively, or to a pair of parallel layers in which one layer is $n$-type and the other is $p$-type (as is common in double-layer graphene), so that one layer has filling factor $\nu$ and the other $-\nu$.  

This setup has been the subject of a number of theoretical studies during the past two decades.\cite{schliemann_strong_2001, yang_dipolar_2001, joglekar_bias-voltage-induced_2002, abolfath_global_2002,  sheng_phase_2003, narasimhan_Wigner-crystal_1995, roostaei_interaction-driven_2015} Most of these studies have focused on describing the phase diagram of the system, and have restricted their analysis to relatively narrow regimes of the particle density, such that $\nu \sim \mathcal{O}(1)$.  In this paper, on the other hand, the emphasis is on describing the scaling behavior of the chemical potential $\mu$ over wide, parametric regimes of density and interlayer separation.  The corresponding analysis highlights a range of phases and behaviors that can be understood from the perspective of an interacting bosonic system.  Specific predictions are also made for the exciton compressibility, which can be used as a probe of the strength of inter-particle interactions and the degree of spatial correlations.  As discussed below, these predictions can be tested using capacitance measurements.

Importantly, the analysis of this paper is restricted to the case of sufficiently high magnetic fields that the magnetic length $\lb$ is much shorter than the zero-field effective electron Bohr radius $a_B^* = \hbar^2 \kappa/(m_e e^2)$.  (Here, $\kappa$ is the dielectric constant and $m_e$ is the electron band mass.)  This condition guarantees that the size of electron/hole wave functions are determined by the radius of the corresponding cyclotron orbits, and are not strongly modified by the electron-hole attraction.  $\lb \ll a_B^*$ is easily satisfied in materials with light mass (or in graphene, where the mass $m_e$ is effectively zero), and in this limit the zero-field mass $m_e$ becomes irrelevant to the exciton behavior.  The high-field condition also guarantees that the contribution of higher Landau levels is unimportant for the exciton behavior.  Finally, this paper also focuses on the case where excitons are dilute: $\nu \ll 1$.  Under this condition the typical exciton momenta $p$ are small, $|p| \ll \hbar/\lb$, and the dispersion relation of excitons is well approximated by a quadratic dependence with a $B$-dependent mass.\cite{lerner_mott_1980, lozovik_magnetoexciton_1997, lozovik_quasi-two-dimensional_2002}

The remainder of this paper is organized as follows.  The following section, Sec.\ \ref{sec:qualitative}, presents a qualitative discussion of results, and outlines the dependence of the interaction energy on the density and the inter-layer separation.  A schematic phase diagram of the system, which includes Wigner crystal and superfluid phases, is also presented.  Sec.\ \ref{sec:numeric} presents a quantitative derivation of the chemical potential at both small and large $d$.  Numeric results are presented which capture the correct asymptotic behavior of the chemical potential and which give an approximate description of the the crossover.  Specific predictions are made for the interlayer capacitance and for the superfluid-Wigner crystal transition.  I conclude in Sec.\ \ref{sec:conclusion} with a discussion of experimental implications.

\section{Scaling derivation of the chemical potential}
\label{sec:qualitative}

In order to understand the qualitative behavior of a system with a finite concentration of excitons, one can first consider the properties of a single exciton.  Under the assumption $\lb \ll a_B^*$ (discussed above), each exciton is composed of a pair of laterally-aligned electron and hole wavepackets of size $\sim \lb$.  [In the symmetric gauge, the corresponding wavefunction for a single electron or hole with zero angular momentum is $\varphi(r) = \exp(-r^2/4\lb^2)/\sqrt{2 \pi \lb^2}$].  The binding energy of the electron-hole pair is given by $\sim -e^2/\kappa r$, where $r \sim \min\{\lb,d\}$ is the typical distance between the electron and hole.  Below I omit this (density-independent) binding energy term from expressions for the chemical potential in order to focus on the density-dependent terms that contribute to the compressibility.

An isolated electron (or an isolated hole) in a strong magnetic field does not have a finite mass.  One can think that because the kinetic energy has been quenched by Landau quantization, the electron mass is effectively infinite.  Nonetheless, the electron-hole bound state does have a finite effective mass, which can be understood as follows.  When the electron and hole wavepackets are given a small lateral displacement $\delta$ relative to each other, there is a corresponding internal Coulomb energy cost $\e_\text{int} \sim e^2 \delta^2/\kappa r^3$.  In this displaced state, the electron and hole also produce for each other an in-plane electric field $E_\parallel \sim e \delta/\kappa r^3$, and as a consequence there is a finite drift velocity $v_d \sim c E_\parallel/B \sim e^2 \lb^2 \delta/(\kappa \hbar r^3)$ resulting from crossed electric and magnetic fields [see Fig.\ \ref{fig:schematic}(b)].  This drift velocity is in the same direction for both the electron and the hole.  Since $v_d \propto \delta$ and $\e_\text{int} \propto \delta^2$, one can define the exciton mass via the relation $\e_\text{int} \sim m v_d^2$.  This relation gives $m \sim \hbar^2 \kappa  r^3/(e^2 \lb^4)$, where, as before, $r \sim \min\{\lb, d\}$.  Effectively, at small $d$ the magnetic length $\lb$ plays the role of the Bohr radius.\cite{lerner_mott_1980}

This finite mass has important implications for the behavior of the excitons at finite density.  In particular, it implies that, unlike for a Wigner crystal of electrons or holes at small filling factor, a Wigner crystal of excitons can melt at low concentration due to quantum fluctuations.   Alternatively, one can say that because quantum fluctuations of the lateral alignment between electron and hole lead to a drift velocity, the exciton has a finite zero-point motion.  It is this zero-point motion that enables the crossover between a Wigner crystal and a superfluid phase for bilayer excitons at finite interaction strength.  Of particular significance for the discussion below is that the mass has a weak dependence on the interlayer separation $d$ at $d \ll \lb$, but it increases strongly with increasing $d$ at $d \gg \lb$:
\be 
m \sim \frac{\hbar^2 \kappa}{e^2 \lb} \times \begin{cases} 
      1, & d/\lb \ll 1 \\
       (d/\lb)^3, & d/\lb \gg 1 
   \end{cases}.
   \label{eq:massq}
\ee
A more exact expression for the mass was derived in Refs.\ \onlinecite{lozovik_magnetoexciton_1997, yang_dipolar_2001}, and is presented in Sec.\ \ref{sec:numeric}.

The interaction potential between excitons takes the form of a dipole-dipole interaction, which is $V(r) \sim e^2 d^3/\kappa r^3$ at distances $r \gg d$.  In two spatial dimensions, this interaction can be classified as short-ranged, with an effective range $b \sim d^2 m e^2/(\kappa \hbar^2)$ that is equivalent to the s-wave scattering length.  [One can arrive at this estimate for $b$ by equating $V(b)$ with $\hbar^2/(mb^2)$.]  At very short distances $r < \lb$, one should think that the $1/r^3$ divergence of the interaction is truncated, since two excitons cannot approach each other closer than the size $\sim \lb$ of the quantum Hall wavefunction.  One can therefore say that two excitons have a maximum interaction energy of $V_\text{max} \sim e^2 d^2/\kappa \lb^3$ at $d \ll \lb$.  

In the remainder of this section, I discuss the behavior of the chemical potential at finite exciton density and its dependence on the two relevant dimensionless parameters: the interlayer separation $d/\lb$ and the filling factor $\nu$.  For the sake of presentation, I assume a filling factor $\nu \ll 1$ and consider regimes of increasing $d/\lb$.

\subsection{$d \ll \lb$: ``Exciton condensate"}
\label{subsec:condensate}
At $d/\lb \ll 1$, the interaction between excitons is weak because of their small dipole moment, and the system assumes a spatially uniform state.  At strictly zero temperature, this state is a BEC, but at finite temperature and in the thermodynamic limit any 2D system has a condensate fraction of zero due to the presence of low-energy fluctuation modes in the superfluid phase.\cite{hohenberg_existence_1967}  Nonetheless, at small but finite temperature this state is a superfluid with long ranged, power-law correlations in the superfluid phase and a chemical potential that is essentially identical to that of a Bose condensate.

To estimate the chemical potential, one can notice that at $d/\lb \ll 1$ the scattering length $b$ becomes shorter than the wavefunction size $\lb$, and therefore only excitons with separation comparable to $\lb$ have a significant interaction.  The interaction energy per exciton can thus be estimated as 
\be 
\mu \sim V_\text{max} \cdot n \lb^2 \sim e^2 d^2 \nu/(\kappa \lb^3).
\label{eq:mucondq}
\ee 
This expression is equivalent to the energy of an exciton Bose condensate, $n \widetilde{V}_0$, where $\widetilde{V}_0$ is the Fourier-transformed interaction law evaluated in the limit of zero wave vector.  

\begin{figure}[htb]
\centering
\includegraphics[width=0.48 \textwidth]{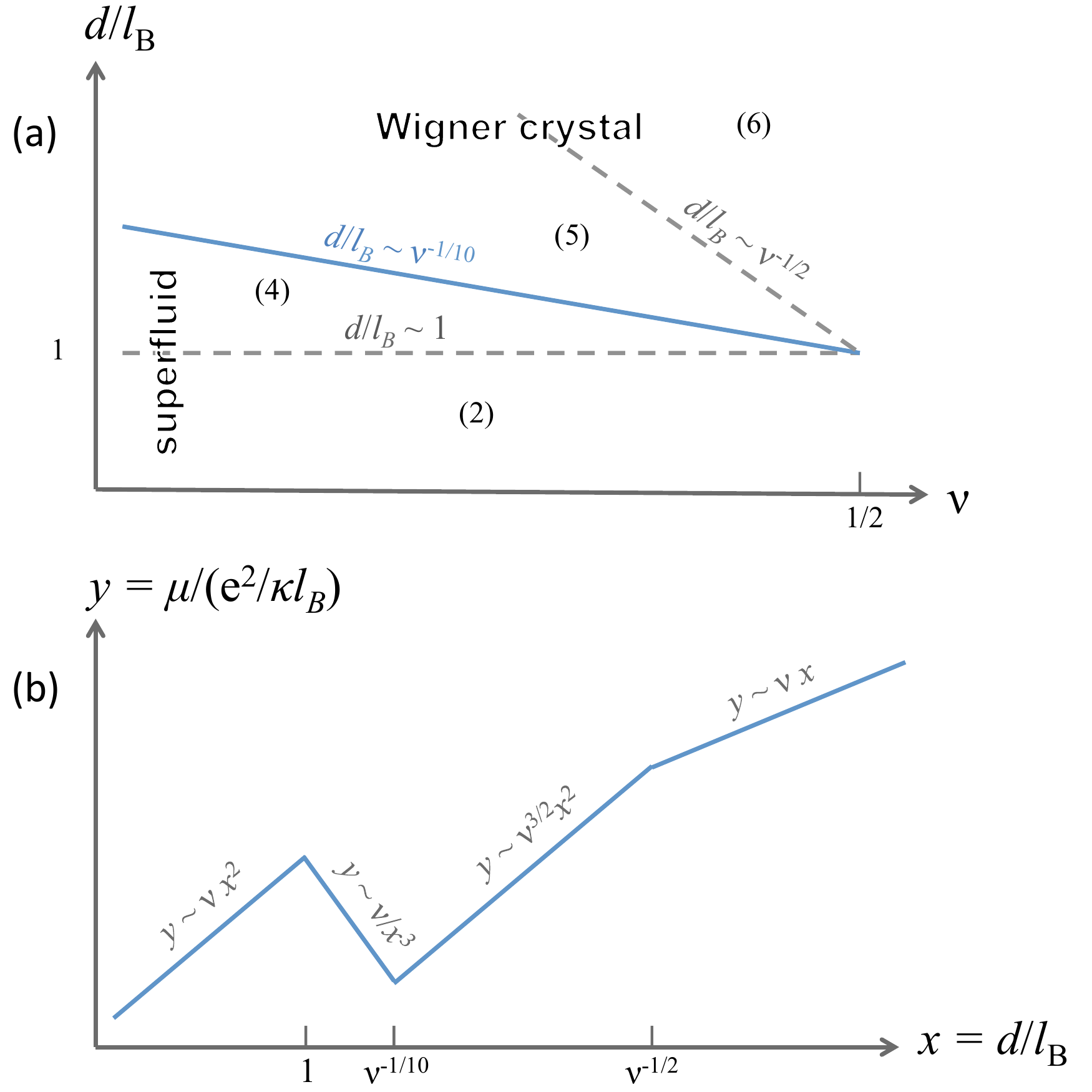}
\caption{(Color online)  Schematic diagram showing the dependence of the chemical potential on filling factor $\nu$ and interlayer separation $d/\lb$.  All axes are depicted in logarithmic scale. (a) Phase diagram showing the superfluid and Wigner crystal phases, as well as the different scaling regimes for the chemical potential $\mu$.  The thick (blue) line denotes a phase transition, and the (grey) dotted lines denote crossovers from one scaling regime to another. Regions of the phase diagram are labeled by the corresponding equation that describes the chemical potential.  The boundaries between different regimes of the phase diagram are labeled by the corresponding equations.   (b) Dependence of the chemical potential on interlayer separation $d$ at a fixed filling factor $\nu \ll 1$.  Different segments of the curve are labeled by the corresponding equation describing them.  For brevity, this plot uses the notation $y = \mu/(e^2/\kappa \lb)$ and $x = d/\lb$. }
\label{fig:diagram}
\end{figure}

The ``exciton condensate" regime is depicted as the bottom region in Fig.\ \ref{fig:diagram}(a).

\

\subsection{$1 \ll d/\lb \ll \nu^{-1/10}$: ``Hard core bosons"} 
\label{subsec:hardcore}
As $d/\lb$ is increased at a given filling factor $\nu \ll 1$, the exciton dipole moment grows and interparticle interactions become stronger.  At $d/\lb \gtrsim 1$ interactions are strong enough that the BEC is destroyed even at zero temperature.  Nonetheless, as long as the typical lateral separation $R = n^{-1/2} \sim \lb \nu^{-1/2}$ between particles is longer than the range $b$ of the interaction, the system remains in a spatially homogeneous phase, which at sufficiently low temperature corresponds to a superfluid.  Equivalently, one can say that the system remains in a spatially homogeneous phase as long as the quantum confinement energy that would be required to spatially separate bosons from each other, $\sim \hbar^2/mR^2$, is larger than the inter-particle interaction energy, $\sim V(R)$.  Inserting the result for the mass at $d/\lb \gg 1$ [Eq.\ (\ref{eq:massq})] into the inequality $\hbar^2/mR^2 \gg V(R)$ gives $\nu \ll (\lb/d)^{10}$, which is equivalent to the condition $b \ll R$.

The chemical potential can be estimated by using the known result for the chemical potential of a 2D gas of hard core bosons with radius $b$,\cite{lieb_ground_2001, schick_two-dimensional_1971}
\be 
\mu \sim \frac{\hbar^2 n}{m \ln\left(\frac{1}{n b^2}\right)}.
\label{eq:muhardcore}
\ee 
That is, in this regime the s-wave scattering length plays the role of an effective ``hard core" radius for the exciton interactions.  Using Eq.\ (\ref{eq:muhardcore}) together with the expression for the effective mass at $d/\lb \gg 1$ gives
\be 
\mu \sim \frac{e^2 \lb^2 \nu} {\kappa d^3 \ln\left(\frac{\lb^{10}}{\nu d^{10}}\right)}.
\label{eq:muhcq}
\ee 

One can notice that Eq.\ (\ref{eq:muhcq}) implies a chemical potential that \emph{decreases} with increasing interlayer separation $d$.  This is a somewhat unusual result, since the strength of interparticle interactions increases monotonically with increasing $d$, and one might therefore expect the energy per particle of a bosonic system to also increase monotonically.  However, the effect of increasing interaction strength is overturned by a sharp growth in the effective mass at $d/\lb > 1$, which reduces the energy associated with quantum confinement.  Consequently, at $d/\lb \gtrsim 1$ the system enters a regime where excitons avoid each other spatially in order to minimize their interaction energy.

I emphasize, however, that the present subsection and the previous one do not describe truly different phases of matter.  As explained in more detail in Sec.\ \ref{sec:numeric}, both regimes correspond to a superfluid, and the transition from Eq.\ (\ref{eq:mucondq}) to Eq.\ (\ref{eq:muhcq}) is merely a crossover associated with renormalization of the interparticle interaction.

\subsection{$\nu^{-1/10} \ll d/\lb \ll \nu^{-1/2}$: ``Dipolar Wigner crystal"}
\label{subsec:dWC}
When $d/\lb$ is increased even further, the exciton-exciton interaction becomes strong enough (and the mass becomes heavy enough) that the system can save energy by undergoing Wigner crystallization.  This excitonic Wigner crystal can be thought of as equivalent to an electron Wigner crystal that is locked with a hole Wigner crystal below it, so that electrons and holes are laterally aligned.  

One can verify that the Wigner crystal description is appropriate by calculating the typical displacement $\delta$ associated with the zero point motion of excitons in the Wigner crystal.  This calculation gives $\delta/R \sim (R \lb^4/d^5)^{1/4} \sim (\lb^{10}/\nu d^{10})^{1/8}$.  The Wigner crystal is stable when $\delta/R \ll 1$, which is equivalent to the condition that the inter-particle spacing $R$ is shorter than the range $b$ of the interaction potential.  This condition is fulfilled when $\nu \gg (\lb/d)^{10}$, which is the opposite limit as in Sec.\ \ref{subsec:hardcore}.  Notice that, unlike for unscreened electron systems, Wigner crystallization corresponds to the limit of \emph{high} density.  This behavior is typical for dipolar systems,\cite{babadi_universal_2013} given the short-ranged interaction, which can overwhelm the kinetic energy only when the interaction energy $\sim e^2 d^2/\kappa R^3$ is much larger than the quantum confinement energy $\sim \hbar^2/mR^2$.

In the Wigner crystal configuration, the dominant contribution to the energy comes from the exciton-exciton interaction at the nearest-neighbor distance: $\mu \sim V(R)$.  As long as the excitons are sufficiently far apart that $R \gg d$, this interaction still has a dipole form, so that 
\be 
\mu \sim \frac{e^2 d^2 \nu^{3/2}}{\kappa \lb^3}.
\label{eq:mudWCq}
\ee

\subsection{$d/\lb \gg \nu^{-1/2}$: ``Normal Wigner crystal"}
\label{subsec:nWC}
Finally, when $d/\lb$ is made so large that the dipole arm $d$ of electron-hole pairs is longer than the interparticle spacing $R$, one can no longer describe interparticle interactions using a $1/r^3$ point dipole approximation.  This regime of $d/\lb \gg R$ corresponds to $\nu \gg (\lb/d)^2$.  

In this regime the energy per particle is dominated by long-ranged Coulomb interactions, and the Coulomb energy of the system is similar to that of two uniformly charged planes.  As in a standard plane capacitor, this energy is $\sim n e^2 d$ per particle.  In other words, at $d/\lb \gg \nu^{-1/2}$, the chemical potential is 
\be 
\mu \sim n e^2 d \sim e^2 d \nu / \lb^2.
\label{eq:munWCq}
\ee 
In-plane positional correlations between excitons give a sub-leading correction to this expression, of order $\sim - e^2 \nu^{1/2}/\kappa \lb$.\citep{bello_density_1981, skinner_anomalously_2010}

Of course, as with Secs.\ \ref{subsec:condensate} and \ref{subsec:hardcore}, the present subsection and the previous one also describe identical phases.  The difference between Eqs.\ (\ref{eq:mudWCq}) and (\ref{eq:munWCq}) is associated only with a crossover in the behavior of the inter-particle interaction law.

The different regimes of behavior for $\mu$ are summarized in Fig.\ \ref{fig:diagram}(a).  The progression of the chemical potential at a fixed filling factor with increasing $d/\lb$ is shown schematically in Fig.\ \ref{fig:diagram}(b).

\section{Analytical expressions and numerical results}
\label{sec:numeric}

In this section I provide a more quantitative derivation of the chemical potential at different values of $\nu$ and $d/\lb$.  These results are then used to calculate the interlayer capacitance and the phase diagram of the system.  I focus first on describing the chemical potential in both the superfluid and Wigner crystal phases, and then present a numerical scheme that reproduces both limits and approximately describes the crossover between them.  Calculations in this section make use of the analytical expressions for the effective mass $m$ and the Fourier-transformed interaction potential $\Vt(q)$ derived in Refs.\ \onlinecite{lozovik_magnetoexciton_1997} and \onlinecite{yang_dipolar_2001}.  These are given by
\be 
\frac{1}{m} = \frac{e^2 \lb}{2 \hbar^2 \kappa} \left[ \sqrt{\frac{\pi}{2}} \left(1 + \frac{d^2}{\lb^2} \right) \exp\left(\frac{d^2}{2\lb^2}\right) \textrm{erfc}\left(\frac{d}{\sqrt{2}\lb}\right) - \frac{d}{\lb} \right]
%\frac{1}{m} = \frac{e^2 \lb}{2 \hbar^2 \kappa} \left[ \sqrt{\frac{\pi}{2}} \left(1 + \frac{d^2}{\lb^2} \right) e^{d^2/2\lb^2} \textrm{erfc}\left(\frac{d}{\sqrt{2}\lb}\right) - \frac{d}{\lb} \right]
\label{eq:m}
\ee 
and 
\be 
\Vt(q) = \frac{4 \pi e^2}{\kappa q} \left( 1 - \exp[-q d] \right) - \Vt_0 ,
%\Vt(q) = \frac{2 \pi e^2}{\kappa q} \left( 1 - \exp[-q d] \right) - \sqrt{8 \pi^3} \frac{e^2 \lb}{\kappa} \left[1 - \exp\left(\frac{d^2}{2\lb^2}\right) \textrm{erfc}\left(\frac{d}{\sqrt{2}\lb}\right) \right].
%\Vt(q) = \frac{2 \pi e^2}{\kappa q} \left( 1 - e^{-q d} \right) - \sqrt{8 \pi^3} \frac{e^2 \lb}{\kappa} \left[1 - e^{d^2/2\lb^2} \textrm{erfc}\left(\frac{d}{\sqrt{2}\lb}\right) \right],
\label{eq:Vq}
\ee
where $\text{erfc}(x)$ is the complementary Gaussian error function.  The first term on the right hand side of Eq.\ (\ref{eq:Vq}) corresponds to the usual dipole-dipole interaction, and the term $-\Vt_0$ represents the truncation of the interaction at short distances $r \lesssim \lb$, with $\Vt_0 = (e^2 \lb/\kappa)\sqrt{8 \pi^3} \left[1 - \exp(d^2/2\lb^2) \textrm{erfc}\left(d/\sqrt{2}\lb\right) \right]$.

\subsection{Superfluid phase}

In a weakly-interacting BEC at zero temperature, the chemical potential is given by $\mu = n \Vt(q=0)$.  While a macroscopic 2D bosonic system at finite temperature cannot strictly be described as a condensate, owing to long-wavelength fluctuations in the phase,\cite{hohenberg_existence_1967} when interactions are weak this expression provides a good approximation to the chemical potential.  As the strength of interactions is increased, however, the condensate-like state is destroyed.  In Ref.\ \onlinecite{fisher_dilute_1988} it was shown that the chemical potential of the superfluid state can described in general by replacing the zero-momentum interaction $\Vt_0 \equiv \Vt(q=0)$ with the value of the self-consistent $t$-matrix.  This procedure gives a self-consistent expression for the chemical potential:
\be 
\mu = \frac{n \Vt_0}{ 1 + (m \Vt_0/4\pi \hbar^2) \ln\left[ \hbar^2/(\mu m a^2 \exp[\gamma_E]) \right] },
\label{eq:muSF}
\ee 
where $\gamma_E \approx 0.5771$ is the Euler-Mascheroni constant and $a$ is the effective interaction range.  The value of $a$ can be taken as the larger of $\lb$ and the s-wave scattering length $b$ for the dipole-dipole interaction, which is given by\cite{meyertholen_biexcitons_2008}
\be 
b = \frac{\hbar^2 \kappa d^2}{m e^2} \exp(2 \gamma_E).
\ee 
A numerical evaluation of Eq.\ (\ref{eq:muSF}) is presented below in Fig.\ \ref{fig:nu01}.

From Eq.\ (\ref{eq:muSF}) one can see the non-monotonic dependence of $\mu$ on the interlayer separation $d$ that was explained qualitatively in Sec.\ \ref{sec:qualitative}.  When $d$ is small, the mass is relatively small and $m\Vt_0/\hbar^2 \ll 1$, and so $\mu \simeq n \Vt_0$, which increases quadratically with $d$.  However, when $d/\lb$ becomes greater than $\sim 1$, the quantity $m \Vt_0/\hbar^2$ becomes larger than unity and dominates the denominator.  Consequently, the chemical potential becomes proportional to $1/m$, which falls as $1/d^3$.  

The critical temperature for the superfluid phase is given by\cite{fisher_dilute_1988}
\be 
T_c \approx \frac{4 \pi \hbar^2 n}{2 m k_B \ln\left[\ln(1/na^2)\right]},
\ee
where $k_B$ is the Boltzmann constant.  At small $d/\lb$, $k_B T_c$ is of order $e^2 \nu/(\kappa \lb)$, while at $d/\lb \gg 1$ it has the smaller magnitude $e^2 \lb^2 \nu/(\kappa d^3)$.

\subsection{Wigner crystal phase}

In dipolar systems the Wigner crystal phase appears at sufficiently high densities that the typical inter-particle spacing is shorter than the interaction range, which here corresponds to $\nu \gg (\lb/d)^{10}$.  In this limit the system is essentially classical, since the zero-point fluctuations in the particle positions are much shorter than the inter-particle spacing, and so the wavefunctions of neighboring excitons have exponentially small overlap.  The dominant contribution to the energy can therefore be found by calculating the electrostatic energy of a classical crystal of dipoles.  This leads to the following expression for the energy $\e$ per exciton:
\be 
\e = \sum_{i,j} \frac{e^2}{\kappa} \left( \frac{1}{r_{ij}} - \frac{1}{\sqrt{r_{ij}^2 + d^2}} \right),
\label{eq:classicalWC}
\ee 
where the indices $\{i,j\}$ label the position of points in a triangular lattice, and run from $-\infty$ to $\infty$, with $\{i,j\} = \{0,0\}$ excluded from the sum.  The variable $r_{ij}$ denotes the Cartesian distance from the origin ($\{i,j\} = \{0,0\}$) to the lattice point $\textbf{r}_{ij}$, and is given by $r_{ij} = \sqrt{2(i^2 + ij + j^2)/(\sqrt{3}n)}$.  The chemical potential is related to the energy per particle by $\mu = d(n \e)/dn$.  Its numeric value is plotted in Fig.\ \ref{fig:nu01} for one particular value of the filling factor $\nu$.

The lowest-order (quantum) correction to Eq.\ (\ref{eq:classicalWC}) can be derived by considering that in a Wigner crystal each particle sits in a parabolic potential well $u(x)$ created by its neighbors, where $x$ indicates a displacement relative to the bottom of the well.  This potential can be calculated by expanding the potential energy created by neighboring particles to second order to give $u(x) \simeq u_0 + \frac12 m \omega^2 x^2$.  The first correction to Eq.\ (\ref{eq:classicalWC}) is then given by $\hbar \omega$, the ground state energy of a 2D harmonic oscillator.

\subsection{Numeric calculations}

Equations (\ref{eq:muSF}) and (\ref{eq:classicalWC}) give relations for the chemical potential that are valid in the asymptotic regimes $d/\lb \ll \nu^{-1/10}$ and $d/\lb \gg \nu^{-1/10}$, respectively.  In order to describe the crossover between these two regimes, one can use a numerical calculation based on the variational principle.  One particularly straightforward choice is to write a variational wavefunction that is a product of Gaussian wave packets $\varphi_{ij}(\rr)$ centered at each point $\rr_{ij}$ of the triangular lattice.  These are given by
\be 
\varphi_{ij}(\rr) = \frac{1}{\sqrt{2 \pi w^2}} \exp \left[- \frac{|\rr - \rr_{ij}|^2}{4 w^2} \right].
\ee 
The width $w$ of the wave packet is used as a variational parameter.  In this state, the kinetic energy per particle is given simply as 
\begin{align}
\e_K & = \int  \varphi_{00}(\rr) \left( -\frac{\hbar^2 \nabla^2}{2 m} \right) \varphi_{00}(\rr) \nonumber  d^2 \rr \\
& = \frac{\hbar^2}{4 m w^2}.
\end{align}
The interaction energy (Hartree energy) per particle can be written as\cite{skinner_effect_2013}
\be
\e_I = \frac{n}{2} \sum_{q \in G} \Vt(q) \exp[-q^2 w^2] - \frac{1}{2} \int_0^\infty dq \frac{q \Vt(q)}{2 \pi} \exp[-q^2 w^2],
\ee 
where $G$ represents the set of all reciprocal lattice vectors of the triangular lattice, and the second term on the right-hand side comes from removing the self-interaction term.  The sum over $G$ amounts to summing over the values of $q$ given by
\be 
q_{ij} = 2 \pi \sqrt{\frac{2 n}{\sqrt{3}}(i^2 + i j + j^2)},
\ee 
where the indices $i, j$ range from $-\infty$ to $\infty$.  The best estimate for the variational parameter $w$ is the one which minimizes the quantity $\e_K + \e_I$.  In the Wigner crystal regime, the optimal value of $w$ is generally such that $w \ll n^{-1/2}$.  In the superfluid regime, on the other hand, $w \gg n^{-1/2}$ and the exciton density is essentially uniform spatially.

\begin{figure}[htb]
\centering
\includegraphics[width=0.5 \textwidth]{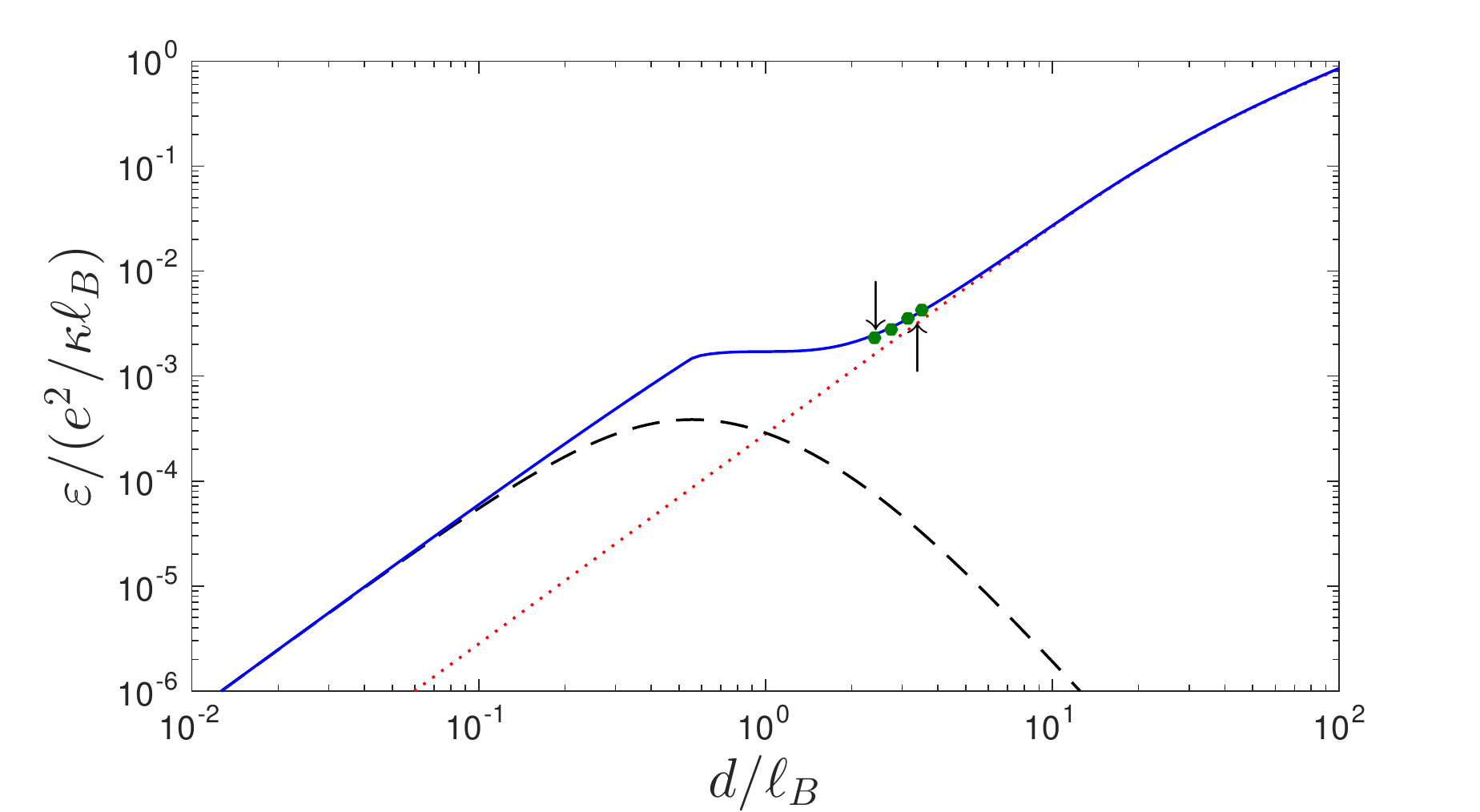}
\caption{(Color online) Dependence of the energy $\e$ per exciton on the inter-layer separation $d$, evaluated at fixed filling factor $\nu = 0.01$.  The (black) dashed line shows the result for the superfluid state given in Eq.\ (\ref{eq:muSF}), and the (red) dotted line shows the energy of a classical Wigner crystal, Eq.\ (\ref{eq:classicalWC}).  The (blue) solid line gives the result from the variational calculation.  The (green) points show results from a QMC study of bosonic point dipoles (Ref.\ \onlinecite{astrakharchik_quantum_2007}), which have been down-sampled for visual clarity.  The down-facing arrow shows the estimate for the liquid-solid transition obtained from the variational calculation, and the up-facing arrow shows the estimate from the QMC studies.}
\label{fig:nu01}
\end{figure}

Figure \ref{fig:nu01} shows the results of this variational calculation at a fixed small filling factor $\nu = 0.01$,  along with the analytical results from Eqs.\ (\ref{eq:muSF}) and (\ref{eq:classicalWC}).  For comparison, I also plot results from a quantum Monte Carlo (QMC) study\citep{astrakharchik_quantum_2007} of bosonic dipoles interacting with the point dipole potential $V(r) = D^2/r^3$, where $D$ is the dipole moment.  Results have been scaled by the corresponding unit of energy $\hbar^6/(m^3 D^4)$, using $D = ed$ and the mass given by Eq.\ (\ref{eq:m}).  While the point-dipole interaction studied in Ref.\ \onlinecite{astrakharchik_quantum_2007} is not identical to the case being studied here, for which excitons have a finite dipole arm $d$, in the limit of $\nu d^2/\lb^2 \ll 1$ the results should be similar.  Indeed, Fig.\ \ref{fig:nu01} shows that the calculated energy is within a few percent of the values from QMC.   

One can notice that in Fig.\ \ref{fig:nu01} the regime of decreasing energy with increasing $d$ depicted in Fig.\ \ref{fig:diagram} is practically not seen due to the narrowness of its window of applicability: $1 \ll d/\lb \ll \nu^{-1/10}$.  Even at the small filling factor $\nu = 0.01$, this window comprises only a factor $\sim 1.6$, and is therefore easily washed out by the crossover behavior.  Still, some indication of this regime is visible in the wide ``shoulder" feature centered at $d/\lb \approx 1$.  A more obvious non-monotonic dependence of $\mu$ on $d$ appears in the numeric solution if $\nu$ is made as small as $0.001$ or smaller.  

\begin{figure}[htb]
\centering
\includegraphics[width=0.5 \textwidth]{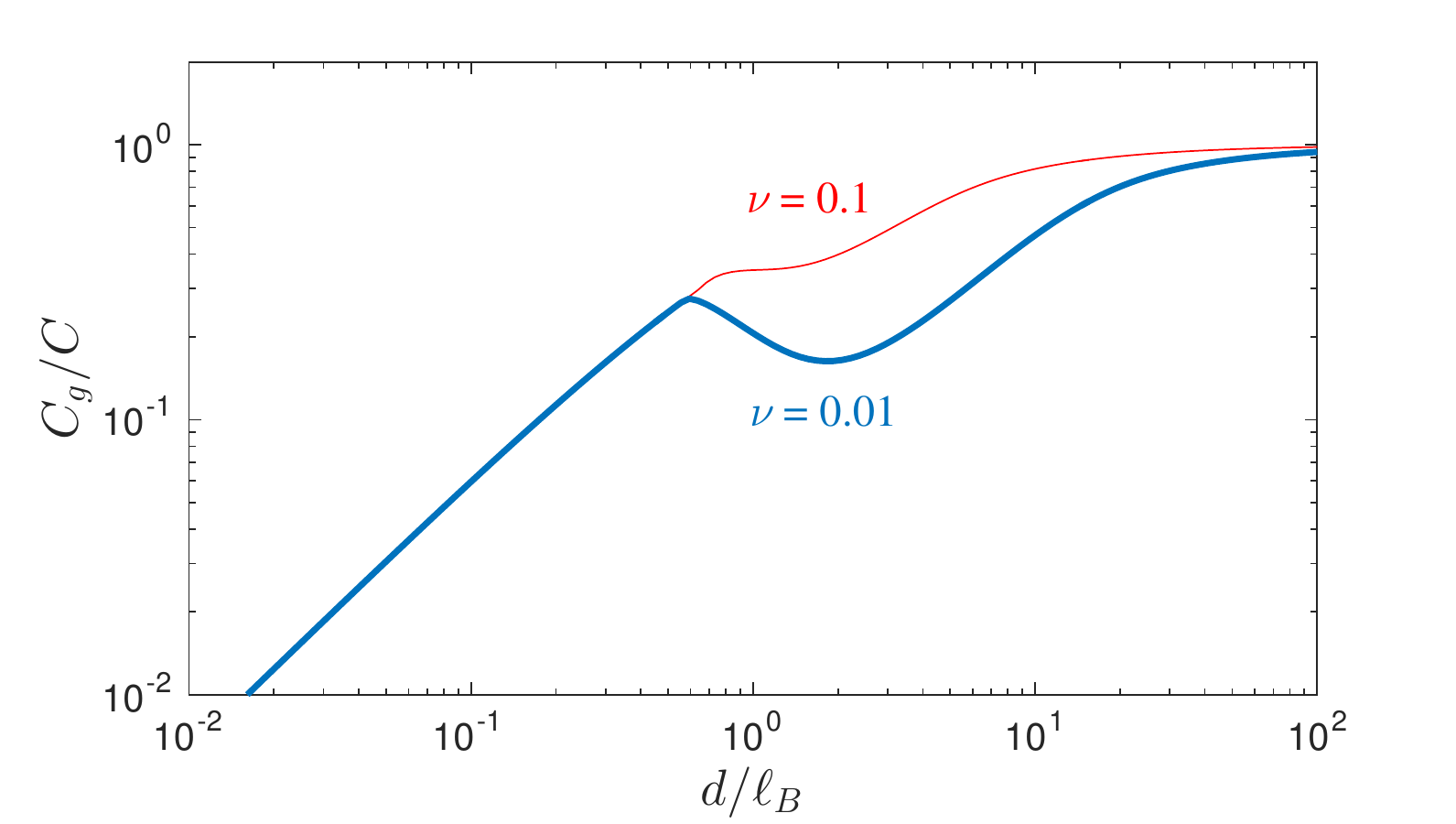}
\caption{Plot of the inverse interlayer capacitance, $C_g/C$, as a function of the interlayer separation $d$.  The curves are labeled by the corresponding value of the filling factor $\nu$.  The local minimum in $C_g/C$ appears near the point of the superfluid-Wigner crystal phase transition.}
\label{fig:C}
\end{figure}

Nonmonotonic behavior is more apparent if one examines the interlayer capacitance $C$ per unit area, defined by $C = e^2 (d\mu/dn)^{-1}$.  As an experimental quantity, the capacitance can be measured using a capacitance bridge,\cite{li_very_2011, skinner_effect_2013} or by field penetration experiments.\cite{eisenstein_negative_1992, eisenstein_compressibility_1994} For a uniform system with large density of states and unscreened Coulomb interactions, the capacitance is given by the ``geometric capacitance" $C_g = \kappa/4 \pi d$; in the present system this situation corresponds to $d/\lb \gg \nu^{-1/2}$.  As $d/\lb$ is reduced, however, the capacitance becomes enhanced over the geometric value due to the presence of strong correlations and the finite range of the dipole-dipole interaction.  This enhancement is shown in Fig.\ \ref{fig:C}, which plots the inverse capacitance $C_g/C$ as a function of $d/\lb$.  Notably, the capacitance acquires a local maximum (and $C_g/C$ acquires a local minimum) near the point of the superfluid-Wigner crystal phase transition.  This maximum is more prominent at small $\nu$.

Finally, one can also use the variational method to produce a quantitative estimate for the superfluid-Wigner crystal phase transition point.  According to the Lindemann criterion of melting, this transition should occur when the uncertainty in the particle position exceeds a critical fraction $\eta_c$ of the lattice constant.  For 2D systems, $\eta_c \approx 0.23$.\cite{astrakharchik_quantum_2007, babadi_universal_2013}  The uncertainty in the particle position $\sqrt{\langle r^2 \rangle }$ is given for the variational wavefunction by $\sqrt{2} w$, so the Lindemann criterion becomes $w = \eta_c n^{-1/2}/3^{1/4}$.  

\begin{figure}[htb]
\centering
\includegraphics[width=0.5 \textwidth]{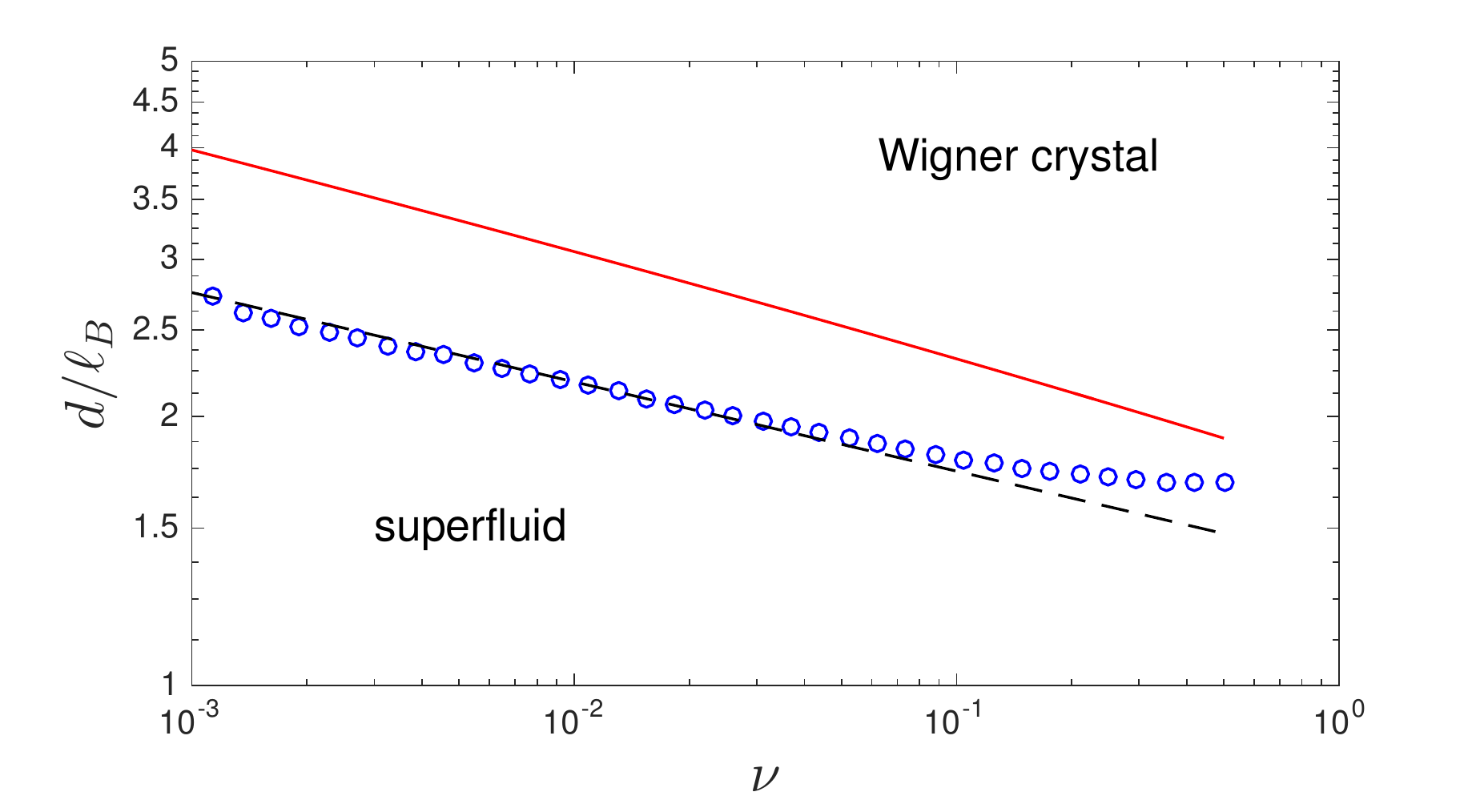}
\caption{(Color online) Phase diagram showing the boundary between the superfluid and Wigner crystal phases, in double-logarithmic scale.  The (red) solid line shows the estimate for the phase boundary obtained by scaling the QMC result from Ref.\ \onlinecite{astrakharchik_quantum_2007}, and the (blue) symbols show results from the numeric solution using the Lindemann criterion.  The dashed line shows a fit to the form $d/\lb \propto \nu^{-1/10}$. }
\label{fig:phasediagram}
\end{figure}

Figure \ref{fig:phasediagram} shows the resulting estimate for the phase boundary.  For comparison, I also plot the QMC result from Ref.\ \onlinecite{astrakharchik_quantum_2007}, which corresponds to $n m^2 e^4 d^4/\kappa^2 \hbar^4 = 290$.  Both are seen to be in good agreement with the analytical prediction $(d/\lb) \propto \nu^{-1/10}$ at small filling factor.

\section{Discussion}
\label{sec:conclusion}

This paper has presented results for the zero-temperature chemical potential of a system of excitons made from electrons and holes in a bilayer quantum Hall system.  One of the primary results is the non-monotonic dependence of the chemical potential on the bilayer separation $d$.  This non-monotonicity arises from a sharp increase in the exciton effective mass when $d$ becomes larger than $\lb$, which leads to a decrease in the renormalized interaction strength and eventually to Wigner crystallization.

One can notice that this non-monotonicity is somewhat difficult to observe experimentally, since it relies on a large separation between $d/\lb = 1$ and $d/\lb = \nu^{-1/10}$.  Nonetheless, the ``shoulder" in $\mu$ that can be seen in Fig.\ \ref{fig:nu01} may be readily observable, and is a sign of the change in inter-particle correlation strength driven by decreasing effective mass with decreasing $d/\lb$.  Probes of the capacitance also exhibit a more prominent non-monotonic behavior.  The existence of a local maximum in the capacitance at $d/\lb \sim \nu^{-1/10}$ may be used as an indication of the Wigner crystal-superfluid phase transition.  

This behavior should be contrasted with that of a single quantum Hall layer adjacent to a metal gate.  Such a system is superficially similar to the one considered here, since screening by image charges in the gate converts the electron-electron interaction into a dipole-dipole interaction.\cite{skinner_anomalously_2010, skinner_simple_2010}  However, in such a screened single-layer system the electrons do not have a finite mass in the limit of high magnetic field, since there is essentially no possibility of an electron being laterally displaced from its image charge in the metal gate.\footnote{In principle, some very large finite mass is possible for an electron-image charge dipole in a high magnetic field, due to the finite response time of the dielectric response in the metal.\cite{sols_bulk_1987}  But this finite time is related to the inverse plasma frequency, which is generally very short.  I therefore expect that any effective mass in the gate-screened single-layer system will be extremely large.}  This leads to a large, qualitative difference in the behavior of the two systems.  For a single-layer system with screening by a metal gate, the electrons remain in a Wigner crystal state down to arbitrarily small values of $d$ and $\lb$, and in this limit their chemical potential has a monotonic dependence on layer separation: $\mu \sim e^2 d^2 \nu^{3/2}/\kappa \lb$.\cite{skinner_giant_2013}  For bilayer excitons, on the other hand, the finite mass implies that the Wigner crystal state melts at $d/\lb \sim \nu^{-1/10}$, and in the limit of small density or small $d/\lb$ the chemical potential is a factor of $\sim \nu^{-1/2}$ larger: $\mu \sim e^2 d^2 \nu/\kappa \lb$.  

Finally, I note that while I have described the transition from Wigner crystal to superfluid using the simple language of ``melting", in principle this phase change is not a simple first-order transition.  For a dipolar system, such a phase change has been shown theoretically to occur through a sequence of ``microemulsion" phases, in which the solid and liquid phases coexist in spatially-mixed stripe or bubble patterns.\cite{spivak_phases_2004}  However, the range of parameter space occupied by these microemulsion phases is related to the discontinuity in density between the solid and liquid phases, which is generally very small.  To my knowledge, such phases have not yet been observed.

\

\acknowledgments 

I am indebted to S.\ Gopalakrishnan, B.\ I.\ Shklovskii, and K.\ Yang for valuable discussions that contributed key ideas to this work.  The first part of this work, comprising the scaling derivation of main results, was completed at Argonne National Laboratory and supported by the U.S. Department of Energy, Office of Basic Energy Sciences under contract number DE-AC02-06CH11357. More exact analytical and numerical calculations were completed at MIT and supported as part of the Center for Excitonics, an Energy Frontier Research Center funded by the U.S. Department of Energy, Office of Science, Basic Energy Sciences under Award no. DE-SC0001088.

%I am indebted to S.\ Gopalakrishnan, B.\ I.\ Shklovskii, and K.\ Yang for valuable discussions that contributed key ideas to this work.
%Work at Argonne National Laboratory was supported by the U.S. Department of Energy, Office of Basic Energy Sciences under contract no.\ DE-AC02-06CH11357.  Work at MIT was supported as part of the Center for Excitonics, an Energy Frontier Research Center funded by the U.S. Department of Energy, Office of Science, Basic Energy Sciences under Award no.\ DE-SC0001088.

\bibliography{qHexcitons}

\end{document}